# Evaluating transition-metal oxides within DFT-SCAN and SCAN+$U$ frameworks for solar thermochemical applications

[1]Gopalakrishnan Sai Gautam and [2]Emily A. Carter

[1]Department of Mechanical and Aerospace Engineering and [2]School of Engineering and Applied Science, Princeton University, Princeton, NJ 08544-5263, United States

## Abstract

Using the strongly constrained and appropriately normed (SCAN) and SCAN+$U$ approximations for describing electron exchange-correlation (XC) within density functional theory, we investigate the oxidation energetics, lattice constants, and electronic structure of binary Ce-, Mn-, and Fe-oxides, which are crucial ingredients for generating renewable fuels using two-step, oxide-based, solar thermochemical reactors. Unlike other common XC functionals, we find that SCAN does not over-bind the $O_2$ molecule, based on direct calculations of its bond energy and robust agreement between calculated formation enthalpies of main group oxides versus experiments. However, in the case of transition-metal oxides (TMOs), SCAN systematically overestimates (*i.e.*, yields too negative) oxidation enthalpies due to remaining self-interaction errors in the description of their ground-state electronic structure. Adding a Hubbard $U$ term to the transition-metal centers, where the magnitude of $U$ is determined from experimental oxidation enthalpies, significantly improves the qualitative agreement and marginally improves the quantitative agreement of SCAN+$U$-calculated electronic structure and lattice parameters, respectively, with experiments. Importantly, SCAN predicts the wrong ground-state structure for a few oxides, namely, $Ce_2O_3$, $Mn_2O_3$, and $Fe_3O_4$, while SCAN+$U$ predicts the right polymorph for all systems considered in this work. Hence, the SCAN+$U$ framework, with an appropriately determined $U$, will be required to accurately describe ground-state properties and yield qualitatively consistent electronic properties for most transition-metal and rare earth oxides.

# Introduction

Generation of reusable fuels or fuel precursors, such as $H_2$, CO, or $CH_4$, using sustainable energy sources, presents an important opportunity to develop carbon-neutral energy storage technologies and sustainable fuels for heavy-duty transportation. Specifically, solar thermochemical (STC) technology could be a crucial component in sustainable fuel (precursor) production, such as in the form of syn-gas (CO+$H_2$), from solar energy, carbon dioxide, and water.[1–4] Typically, a two-step reduction/re-oxidation process involving a redox-active oxide substrate is employed to generate fuel precursors. For the thermal reduction (TR) step, the oxide substrate is heated to high temperatures to induce oxygen off-stoichiometry and subsequent oxygen loss, where the reduction reaction can be written as $\frac{1}{\delta}MO_x \rightarrow \frac{1}{\delta}MO_{x-\delta} + \frac{1}{2}O_2(g)$. For the steam/$CO_2$ gas splitting (GS) step, the reduced oxide is cooled without any re-oxidation to a lower temperature at which the oxide can react exothermically with steam or $CO_2$ to generate $H_2$ or CO, respectively, where the re-oxidation reaction can be written as $\frac{1}{\delta}MO_{x-\delta} + H_2O/CO_2(g) \rightarrow \frac{1}{\delta}MO_x + H_2/CO\ (g)$. STC technologies theoretically can achieve high efficiencies[5,6] because they harvest the entire solar spectrum, in contrast to photovoltaic-aided or photoelectrocatalytic water/$CO_2$ splitting that only captures those photons with energies larger than the material's band gap. However, the viability of STC reactors depends heavily on the oxide substrate used.[7] Specifically, the oxide must be thermally stable across a wide range of temperatures and able to generate large amounts of desired products. A quantum-mechanics-based search for potential candidates[8–12] could accelerate the design and development of STC reactor materials, given that prior density functional theory (DFT)[13,14] based searches have yielded several successful candidate materials for other energy (and allied) applications.[15,16]

Materials that have been considered thus far for STC applications belong to three structural categories: *i*) $AO_2$ compounds, such as pure and Zr-doped $CeO_2$,[17–19] which adopt the fluorite structure and have an oxygen:metal ratio of 2:1; *ii*) $ABO_3$ compounds, such as (La,Sr)$MnO_3$,[20,21] which exhibit a perovskite structure and have an oxygen:metal ratio of 3:2; and *iii*) $AB_2O_4$ compounds, such as the spinel-Fe(Fe,Al)$_2O_4$,[22,23] with an oxygen:metal ratio of 4:3. Note that at least one metal atom type in the aforementioned compounds must be redox active in order to be a viable candidate for STC applications. For example, redox-active Ce ($Ce^{4+} \leftrightarrow Ce^{3+}$) can facilitate $CeO_2$ to be a viable candidate for both TR and GS. Similarly, $Mn^{2+/3+/4+}$ and $Fe^{2+/3+}$ are the redox-active species in the perovskite and spinel materials, respectively. Thus, any theory-based evaluation of potential STC candidates requires a rigorous, accurate description of reduction and oxidation energetics amongst transition-metal oxides (TMOs) and rare-earth oxides (REOs, such as $CeO_2$). Specifically, the choice of the functional describing the electron exchange-correlation (XC) interactions, within the framework of DFT, strongly influences redox energetics. Note also that any level of theory that can adequately describe redox energetics of TMOs will be of significant importance in related fields of photovoltaics, batteries, and photoelectrocatalysts.[24–27]

The strongly constrained and appropriately normed (SCAN) XC functional was developed recently by Perdew and co-workers.[28] SCAN importantly satisfies the 17 known constraints on the behavior of XC functionals, unlike the local density approximation (LDA) or the generalized gradient approximation (GGA).[28–30] Calculations using SCAN, so far, indicate that SCAN accurately predicts formation energies of main group (*i.e.*, *s* and *p*) oxides[31–33] and sulfides.[34] SCAN also predicts the right polymorph stability in select TMOs, such as $MnO_2$.[35] However, it remains to be seen if SCAN can predict the energies of redox reactions involving TMOs.

Our previous work on defects in $Cu_2ZnSnS_4$-based solar cells[34] indicated that SCAN significantly underestimates the band gap of transition-metal-containing semiconductors, analogous to the behavior of the GGA XC functional.[36,37] Such underestimation of band gaps usually leads to an erroneous description of the electronic ground state, and, as a result, an erroneous ground-state energy.[38,39] Thus, redox processes, which typically involve electron transfer across significantly different electronic environments (say, from a metal to an insulator or from an oxygen *p* to a metal *d* orbital), are likely to be erroneously described by SCAN. Errors in redox energy predictions from SCAN nominally are expected to be particularly severe in highly ionic environments, such as *d* and *f* oxides, with significant electronic exchange and correlation (*i.e.*, amongst *d* and *f* electrons).[40–43] Importantly, *d* (and *f*) orbitals are more localized than analogous *s* and *p* orbitals, leading to stronger XC interactions between the electrons.

Shortcomings of GGA XC functionals, such as poor descriptions of redox energetics and band gaps, have been overcome by the addition of a Hubbard $U$ term,[44] resulting in a GGA+$U$ functional. Typically, the $U$, formulated as $U_{eff} = U\text{-}J$, is added to the transition-metal (TM) atoms that contain the $d$ electrons, as a penalty term that accounts for the on-site Coulomb ($U$) and exchange ($J$) interactions. We will refer to $U_{eff}$ simply as $U$ henceforth. However, the magnitude of $U$ for each TM atom is not known *a priori* and is normally dependent on the choice of the XC functional, which itself is a source of error.[45–49] The value of $U$ for a given oxidation state of a TM (say, $Fe^{2+}$) instead can be determined independently from first principles, based on electrostatically embedded Hartree-Fock calculations as originally developed by Mosey *et al.*,[40,41] but at significant computational expense. Alternatively, the magnitude of $U$ can be fitted either to measured oxidation (or equivalently reduction) energies[42,50] or to measured band gaps,[45] with the caveat that the latter is not well-founded since DFT+$U$ eigenvalue gaps are not actual experimental observables. Although fitting the $U$ to experimental quantities is computationally inexpensive, subsequent predictions may not be as transferable as desired, *i.e.*, calculations with a $U$ fitted to oxidation energies do not necessarily predict band gaps accurately and vice-versa. Determining $U$ from oxidation energies nevertheless yields an "average" $U$ value across the oxidation states considered, which permits subsequent unbiased calculations for redox reactions within solids.[43] For example, an average $U$ value determined for the $Fe^{2+} \leftrightarrow Fe^{3+}$ redox can be used to calculate the spinel-Fe(Fe,Al)$_2$O$_4$ at various oxygen stoichiometries (*i.e.*, Fe(Fe,Al)$_2$O$_{4-\delta}$) without prior identification of the specific Fe atoms retaining a +2 or +3 oxidation state within the structure. Indeed, calculations based on average $U$ values have been ubiquitous in describing redox-active TMOs that are normally used as electrode materials within batteries,[51–54] often with good agreement between predicted and measured voltages.

In this work, we consider the oxidation energetics of binary TMOs, specifically Ce-, Mn-, and Fe-oxides, which are critical for a reliable theoretical description of candidate and benchmark materials for STC applications. Given that the GGA is known to over-bind the O$_2$ molecule, causing underestimation of oxidation energies (*i.e.*, DFT-GGA oxidation energies are less negative than experimental values),[42,55,56] we initially checked if SCAN exhibits similar trends, by calculating the O$_2$ bond dissociation energy and formation energies of main group oxides (without any *d* or *f* electrons). We find that DFT-SCAN formation energies display robust agreement with experimental values of both sets of properties.

Subsequently, we determine average *U* values across the possible oxidation states of Ce (+3/+4), Mn (+2/+3/+4), and Fe (+2/+3) based on experimental oxidation energies. We then compare experimental lattice constants, band gaps, and TM magnetic moments of binary oxides containing these ions using both DFT-SCAN and SCAN+*U*, where we simply use total density of states (DOS) calculations to estimate band gaps. We also highlight important qualitative trends that emerge from our calculations on these oxides, which can explain the trends observed in oxidation energetics. Specifically, we analyze *i*) the polymorph predicted as the ground state for $Ce_2O_3$ by DFT-SCAN and SCAN+*U*; *ii*) various magnetic configurations that are stabilized in $Mn_2O_3$; and *iii*) the variation of the electronic structure in $Fe_3O_4$. Finally, we strongly suggest using the SCAN+*U* XC functional (with an appropriately determined *U*) for theoretical studies involving redox reactions of all TMOs and REOs, with potential extensions to ionic sulfide compounds, as well.[24,57]

## Methods

All calculations are performed spin-polarized using the Vienna ab initio simulation package (VASP),[58,59] employing the all-electron, frozen-core projector-augmented-wave (PAW) theory.[60] We use SCAN for describing the XC of all metals, oxygen, and main group oxides (see below), while we use both SCAN and SCAN+*U* for binary Ce-, Mn-, and Fe-oxides. For performing SCAN+*U* calculations, we employ the simplified rotationally invariant framework developed by Dudarev *et al*.[61] We describe the electronic one-electron wavefunctions with a plane-wave basis, up to a kinetic energy cutoff of 520 eV, and sample them with a dense Γ-centered *k*-point grid (with a spacing of ~0.03 Å$^{-1}$), which converges total energies to within ~1 meV/atom (convergence behavior indicated in **Figure S1** of the Supporting Information (SI)[62]). While relaxing a given structure, we converge the total energies and the atomic forces up to < 0.01 meV and < |0.03| eV/Å, respectively, within that structure. For all oxides and metals, except MnO and FeO (see description of magnetic configurations below), we use the conventional unit cell for all calculations, as obtained from the inorganic crystal structure database (ICSD).[63] For calculating the oxygen molecule and an isolated oxygen atom, we use an asymmetric 15 Å×16 Å×17 Å cell to obtain the appropriate triplet spin ground states of both $O_2$ and O. We have included sample VASP INCAR input files and a list of PAW potentials used in our calculations in the SI.

**Reaction energies**

For determining average $U$ values, we utilize oxidation energies of binary Ce-, Mn-, and Fe-oxides. Schematically, the oxidation reaction for the aforementioned oxides can be written, normalized per mole of $O_2$, as $MO_x + \frac{z-x}{2} O_2 \rightarrow MO_z$. Based on experimentally tabulated standard formation enthalpies (at 298 K and 1 atm) of $MO_x$ and $MO_z$, available via the Kubaschewski and Wagman tables,[64,65] we can estimate the experimental oxidation enthalpy per mole of $O_2$ as $\Delta H_o^e = \frac{H^0_{MO_z} - H^0_{MO_x}}{\frac{z-x}{2}}$. Note that we use experimental values at 298 K and 1 atm, since experimental 0 K data are not uniformly available across all systems. Analogously, the theoretically predicted oxidation enthalpy can be written based on 0 K energies ($E$), while neglecting the $P\Delta V$ contribution, as $\Delta H_o^t = \frac{E^{SCAN+U}_{MO_z} - E^{SCAN+U}_{MO_x} - \frac{z-x}{2} E^{SCAN}_{O_2}}{\frac{z-x}{2}}$ (see discussion on the sensitivity of $P\Delta V$ and zero-point energy ($ZPE$) contributions on theoretical enthalpy calculations in the SI; $P\Delta V + ZPE$ is only included for calculating the $O_2$ bond dissociation energy). Note that $U = 0$ in the expression for $\Delta H^t$ indicates a SCAN calculation without any $U$ term added. Finally, we estimate the $U$ value so as to minimize the absolute error between $\Delta H_o^e$ and $\Delta H_o^t$. In the case of main group oxides, we compare the experimental and theoretical formation energy ($\Delta H_f$) of a given oxide. For example, the theoretical formation energy for the formation of MgO from Mg is written as $\Delta H_f^t = \frac{E^{SCAN}_{MgO} - E^{SCAN}_{Mg} - \frac{1}{2} E^{SCAN}_{O_2}}{\frac{1}{2}}$, while the experimental formation energy is tabulated.[64,65]

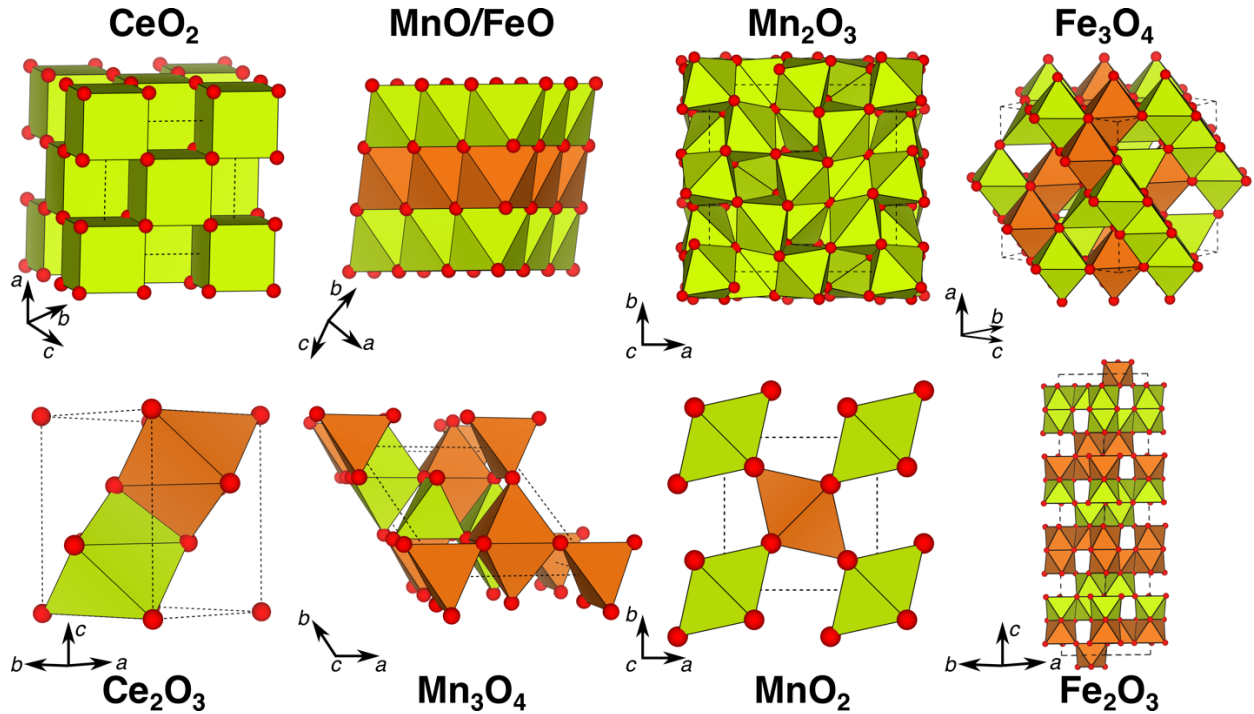

**Figure 1:** Crystal structures of all of the transition-metal oxides considered in this work. Yellow and orange polyhedral indicate TM atoms adopting up and down magnetic moments, respectively. The crystal structures displayed here correspond to the ground-state configuration within the SCAN+$U$ framework at the optimal $U$ value determined in this work. Table 1 lists the specific space groups of all structures.

**Crystal structures**

**Figure 1** displays the crystal structures of all Ce-, Mn-, and Fe-oxides considered in this work. In addition to these polymorphs, we included all "ordered" crystal structures, *i.e.*, structures where occupancies of all atomic sites equal an integer, that are available in the ICSD for each composition. Notably, all polymorphs in **Figure 1** correspond to the ground-state configuration within the SCAN+$U$ framework at the determined optimal $U$ value; **Table 1** lists their space groups. For each structure, we calculated the energies of both ferromagnetic (FM) and specific antiferromagnetic (AFM) configurations. The yellow (orange) polyhedra of each ground-state configuration in **Figure 1** correspond to the TM atom within that polyhedron adopting an up (down) magnetic moment. To calculate the formation energies of the main group oxides, we used *i*) rocksalt (space group: $Fm\bar{3}m$) MgO, CaO; *ii*) anti-fluorite ($Fm\bar{3}m$) Li$_2$O, Na$_2$O, and K$_2$O; *iii*) hexagonal ($P6_3mc$) BeO; *iv*) corundum ($R\bar{3}c$) $\alpha$-Al$_2$O$_3$; and *v*) quartz ($P3_121$) $\alpha$-SiO$_2$. For calculating the corresponding pure elements, we employed *i*) hexagonal-close-packed ($P6_3/mmc$) Be, Mg; *ii*) face-centered-cubic ($Fm\bar{3}m$) Ca, Al; *iii*) body-centered-cubic ($Im\bar{3}m$) Li, Na, and K; and *iv*) diamond-cubic ($Fd\bar{3}m$) Si.

**Magnetic configurations**

To capture the type-II antiferromagnetism of MnO and FeO,[66] we employed a $2 \times 2 \times 2$ supercell of the primitive rocksalt structure. Spinel-Fe$_3$O$_4$ (space group: $Fd\bar{3}m$) exhibits a significant degree of "inversion"[67,68] and electronic conductivity[69] at room temperature. However, Fe$_3$O$_4$ undergoes a Verwey transition at low temperatures (~120 K[70,71]), which leads to a small opening of the band gap, a slight distortion from the spinel to an orthorhombic structure, and a Fe$^{2+}$-Fe$^{3+}$ charge ordering on the octahedral sites.[69,72,73] Thus, we used the AFM (ferrimagnetic) model proposed by Wright *et al.*,[69] without preserving the symmetry in our structural relaxation calculations to facilitate any distortions away from the spinel structure. In the case of the tetragonally distorted spinel-Mn$_3$O$_4$ (space group: $I4_1/amd$), which contains Jahn-Teller active Mn$^{3+}$ ions, the magnetic ground-state configuration is not known unequivocally.[74–76] Previous studies, such as by Chartier *et al.*,[75] considered six different ferrimagnetic configurations, using the notation of "FIM$_x$", where x is the number assigned to a configuration. Thus, we calculated the energies of all six ferrimagnetic configurations of Mn$_3$O$_4$ at both DFT-SCAN and SCAN+$U$ levels of theory and found the "FIM$_6$" configuration to be the most stable in both electronic structure frameworks (**Figure 1**).

Similarly, the ground-state magnetic configuration of $\alpha$-Mn$_2$O$_3$ (space group: $Pbca$) is still debated.[76,77] Given the large unit cell required to describe the structure of $\alpha$-Mn$_2$O$_3$, it is computationally prohibitive to explore all possible AFM orderings. Hence, we considered the AFM ordering proposed by Regulski *et al.*[77] in addition to the FM configuration. For Fe$_2$O$_3$,[78,79] Ce$_2$O$_3$,[80,81] and MnO$_2$,[76,82,83] we used spin-models identical to those used to describe the corresponding magnetic properties. For example, the two unique metal atoms within the conventional unit cells of Ce$_2$O$_3$ and MnO$_2$ are set to equally opposite magnetic moments to represent the AFM configuration. In the case of Fe$_2$O$_3$, the magnetic moment changes sign across every two layers of Fe-atoms along the *c*-direction (**Figure 1**). Also, we initialized the electronic spin state of each TM atom in all of the structures considered in their corresponding high-spin (HS) state, motivated by prior experimental evidence.[80,83–89] For example, each $Fe^{3+}$ atom in the AFM-$Fe_2O_3$ structure was initialized in the HS $d^5$ configuration, where each *d* electron singly occupies individual *d* orbitals.

## Results

**Formation energies of main group oxides**

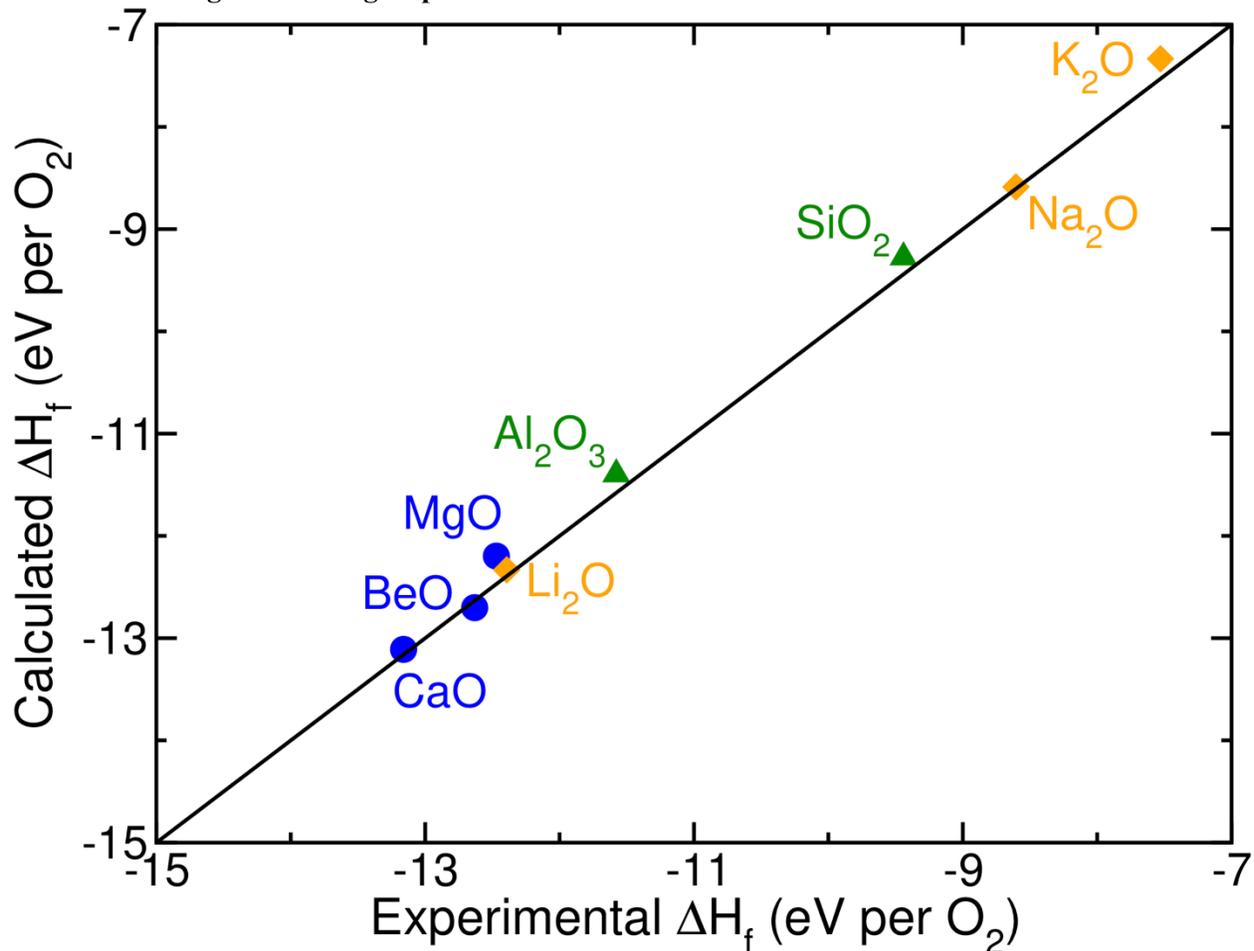

**Figure 2:** DFT-SCAN formation energies of main group (*s* and *p*) oxides plotted against experimental formation energies.[64,65] Yellow diamonds, blue circles, and green triangles indicate oxides of alkali, alkaline earth, and *p*-block elements, respectively. Solid black line signifies parity between theoretical predictions and experiments.

**Figure 2** plots the formation energies of main group oxides as calculated by DFT-SCAN, against experimental data. Oxides of alkali, alkaline earth, and *p*-block elements are respectively indicated by yellow diamonds, blue circles, and green triangles, respectively, while the solid black line corresponds to the parity between theoretical predictions and experimental values. Note that we consciously considered main group elements that did not contain any closed-shell *d* or *f* orbitals (*e.g.*, Ga, Sn, Bi, *etc.*) to efficiently isolate potential errors in oxide formation energies due to over-binding (or under-binding) of the oxygen molecule alone. The data in **Figure 2** indicate excellent agreement between DFT-SCAN and experimental formation energies of all main group oxides, with a mean absolute error (MAE) of ~0.1 eV per $O_2$. The largest deviation in DFT-SCAN predictions versus experiments is for rocksalt MgO, where DFT-SCAN underestimates the formation energy by ~0.2 eV per $O_2$ (~2% error). Analogous comparisons of DFT-GGA formation energies versus experiments previously yielded a consistent underestimation of ~1.36 eV per $O_2$.[42] Note that our DFT-SCAN calculations predict a bond length of 1.22 Å for an isolated $O_2$ molecule, in close agreement with previous DFT-GGA calculations (~1.22 Å[55]) and experiment (~1.21 Å). Additionally, the DFT-SCAN $O_2$ bond dissociation energy (~5.15 eV) is in excellent agreement with experiments (~5.12 - 5.23 eV[90,91]), in contrast to DFT-LDA (~7.2 - 7.6 eV[55,91]) and DFT-GGA (~5.7 - ~6.2 eV[42,55,56]). SCAN thus does not cause any errors in oxide formation energies due to its excellent description of the oxygen molecule's ground state (unlike GGA) and therefore may be highly suitable for describing the energetics of main group elements and compounds.[31,34]

**Oxidation energetics of TMOs**

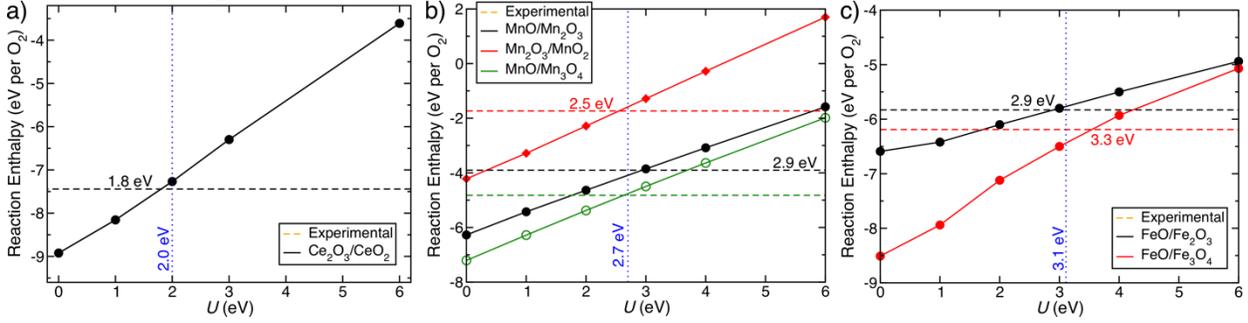

**Figure 3:** Variation of the oxidation reaction enthalpy (solid lines) in Ce-oxides (panel a), Mn-oxides (b), and Fe-oxides (c), with increasing magnitude of $U$ within the SCAN+$U$ framework. Horizontal dashed lines in each panel reflect experimental oxidation enthalpies[64,65] for each reaction considered, with the text annotations adjacent to each dashed line signifying the $U$ value that minimizes error between experimental data and SCAN+$U$ predictions for the corresponding reaction. Blue dotted lines and text indicate optimal $U$ values, obtained by averaging the $U$ values for individual oxidation reactions. In the case of Mn-oxides (b), the MnO→$Mn_3O_4$ oxidation reaction deliberately was not used in determining the optimal $U$.

**Figure 3** plots the enthalpy of oxidation reactions (solid lines) within Ce-oxides ($Ce_2O_3$ → $CeO_2$, panel a), Mn-oxides (panel b), and Fe-oxides (panel c), respectively, as a function of $U$ used in SCAN+$U$ calculations. In the case of Mn-oxides, we considered three distinct oxidation reactions, namely MnO → $Mn_2O_3$ (solid black line, **Figure 3b**), $Mn_2O_3$ → $MnO_2$ (solid red line), and MnO → $Mn_3O_4$ (solid green line). Similarly, we evaluated the oxidation reactions of FeO → $Fe_2O_3$ (solid black line in **Figure 3c**) and FeO → $Fe_3O_4$ (solid red line) for Fe-oxides. The dashed lines in each panel correspond to the experimental oxidation enthalpies.[64,65] For example, the dashed red line in **Figure 3b** indicates the enthalpy of $Mn_2O_3$ → $MnO_2$, whose SCAN+$U$ calculated values are signified by the solid red line. Finally, the dotted blue lines in each panel of **Figure 3** reflect the optimal $U$ that minimizes the error between SCAN+$U$ predicted and experimental oxidation enthalpies for the oxidation reactions considered for each system. Notably, we did not consider the MnO → $Mn_3O_4$ oxidation reaction when determining the optimal $U$ for Mn ($U_{Mn}$ = 2.7 eV) to test the transferability of SCAN+$U_{Mn}$ calculations. Also, previous work[25,26,92] has demonstrated the relative insensitivity in energy and band-gap trends for variations of ±0.5 eV in the magnitude of $U$ used.

Predictions of oxidation enthalpies by DFT-SCAN, *i.e.*, $U = 0$ in all panels of **Figure 3**, indicate significant disagreement with experimental measurements. For example, the SCAN-predicted oxidation enthalpy for $Ce_2O_3 \rightarrow CeO_2$ (**Figure 3a**) is -8.92 eV per $O_2$, substantially more negative than the experimental -7.44 eV per $O_2$. The probable cause of this discrepancy in Ce-oxides is a significant underestimation of the ground-state energy of $Ce_2O_3$ by DFT-SCAN, which arises from an inaccurate description of the ground-state electronic structure (see the section on $Ce_2O_3$ below). Similar trends can be observed in Mn- and Fe-oxides as well, with DFT-SCAN oxidation enthalpies substantially more negative than experimental values. For example, DFT-SCAN predicts oxidation enthalpies of -6.27, -4.21, and -7.21 eV per $O_2$ for $MnO \rightarrow Mn_2O_3$, $Mn_2O_3 \rightarrow MnO_2$, and $MnO \rightarrow Mn_3O_4$, respectively (**Figure 3b**), against the experimental -3.90, -1.73, and -4.82 eV per $O_2$. In the case of Fe-oxides, DFT-SCAN (experimental) values are -6.59 (-5.83) and -8.51 (-6.31) eV per $O_2$ for $FeO \rightarrow Fe_2O_3$ and $FeO \rightarrow Fe_3O_4$, respectively. Given the substantial deviations and the consistently more negative (or overestimation) of DFT-SCAN oxidation enthalpies with respect to experiment, across different TMOs, SCAN+$U$ calculations are essential for describing any redox energetics within such materials. Since the magnitude of $U$ is not known *a priori*, we have determined optimal $U$ values for all TM atoms considered in this work based on the corresponding oxidation energies of their binary oxides.

The text annotations in each panel of **Figure 3**, along the dashed lines, indicate the ideal $U$ value that minimizes the absolute error between SCAN+$U$ and experimental enthalpies for the corresponding oxidation reaction. For example, we determine the ideal $U$ value for $Ce_2O_3$ oxidizing to $CeO_2$ to be ~1.8 eV (**Figure 3a**). We find, however, that SCAN+*1.8* stabilizes the wrong ground-state polymorph in $Ce_2O_3$ (see the $Ce_2O_3$ section below), prompting us to use a slightly higher $U_{Ce} = 2$ eV. In the case of Mn-oxides, we find ideal $U$ values for $MnO \rightarrow Mn_2O_3$ and $Mn_2O_3 \rightarrow MnO_2$ to be 2.9 and 2.5 eV, respectively (**Figure 3b**). Thus, the optimal $U$ for Mn oxidation states between +2 and +4 ($U_{Mn}$) is evaluated as the average of the aforementioned oxidation reactions, resulting in a value of 2.7 eV (dotted blue line in **Figure 3b**). Notably, SCAN+*2.7* predicts the oxidation enthalpy for $MnO \rightarrow Mn_3O_4$ (not used in determining $U_{Mn}$) to be -4.76 eV per $O_2$, which is in close agreement with the experimental -4.82 eV per $O_2$, indicating that $U_{Mn}$ can efficiently describe redox energetics involving $Mn^{+2/+3/+4}$ in other structures. Analogously, the ideal $U$ values for FeO → $Fe_2O_3$ and FeO → $Fe_3O_4$ are 2.9 and 3.3 eV, respectively, with the optimal $U$ ($U_{Fe}$) equaling 3.1 eV.

**Lattice parameters and band gaps**

Apart from oxidation energetics, we also have benchmarked the lattice constants, band (eigenvalue) gaps, and TM magnetic moments for all of the TMOs considered, with DFT-SCAN and SCAN+$U$; the results appear in **Table 1**. The space group of each structure considered, pictorially represented in **Figure 1**, is listed below each composition. The experimental data (including lattice parameters,[63] band gaps, and magnetic moments) for Ce-, Mn-, and Fe-oxides are obtained from Refs. [80,93,94], [83–86,95–100], and [73,79,87–89,101], respectively. The $U$ values used for SCAN+$U$ calculations are the optimal $U$ values indicated in **Figure 3**; specifically, $U_{Ce}$ = 2 eV, $U_{Mn}$ = 2.7 eV, and $U_{Fe}$ = 3.1 eV. The magnetic moments ($m$) indicated are in units of Bohr-magneton ($\mu_B$) and reflect the absolute magnetic moments amongst alike TM atoms within each TMO. For example, $m_{Fe}$ in $Fe_2O_3$ corresponds to the absolute magnetic moment amongst all $Fe^{3+}$ ions within AFM-$Fe_2O_3$. In the case of structures such as $Mn_3O_4$ and $Fe_3O_4$, which contain two distinct oxidation states of TM atoms, we list the moments for both oxidation states (see the foot-notes below **Table 1**). The lattice parameters listed for MnO and FeO correspond to a 2×2×2 supercell of the corresponding primitive rocksalt lattices. While spinel-$Fe_3O_4$ undergoes a slight cubic → orthorhombic distortion below the Verwey transition temperature (~120 K),[70] the experimental lattice parameters in **Table 3** correspond to room-temperature values (the experimental values below 120 K are not unambiguously known[69,72]).

**Table 1:** Experimental, DFT-SCAN, and SCAN+$U$ lattice constants, lattice angles, band gaps, and magnetic moments of the corresponding TM atoms for all TMOs considered in this work. Specific $U$ values used are $U_{Ce}$ = 2.0 eV, $U_{Mn}$=2.7 eV, and $U_{Fe}$ = 3.1 eV. The space group of the polymorph evaluated is indicated below each composition.

| Composition (Space Group) | Source | Lattice constants (Å) | | | Lattice angles (°) | | | Band gap (eV) | $m$ ($\mu_B$) |
|---|---|---|---|---|---|---|---|---|---|
| | | $a$ | $b$ | $c$ | α | β | γ | | |
| CeO$_2$ ($Fm\bar{3}m$) | Expt. | | 5.41 | | | 90 | | 6 | - |
| | SCAN | | 5.42 | | | 90 | | 1.79 | 0.0 |
| | SCAN+$U_{Ce}$ | | 5.44 | | | 90 | | 1.93 | 0.0 |
| Ce$_2$O$_3$ ($P\bar{3}m1$) | Expt. | | 3.89 | 6.06 | 90 | | 120 | 2.4 | 1.08 |
| | SCAN | | 3.85 | 6.04 | 90 | | 120 | 0.0 | 0.89 |
| | SCAN+$U_{Ce}$ | | 3.88 | 6.06 | 90 | | 120 | 1.11 | 0.94 |
| MnO ($Fm\bar{3}m$) | Expt. | | 6.29 | | | 60 | | 3.6-3.8 | 4.58 |
| | SCAN | 6.27 | 6.18 | 6.27 | 59.5 | 60 | 59.5 | 0.62 | 4.43 |
| | SCAN+$U_{Mn}$ | 6.31 | 6.24 | 6.31 | 59.6 | 60 | 59.6 | 1.2 | 4.57 |
| Mn$_2$O$_3$ (Pbca) | Expt. | 9.41 | 9.42 | 9.40 | | 90 | | 1.2-1.3 | 3.1-4.2 |
| | SCAN | 9.40 | 9.35 | 9.40 | | 90 | | 0.0 | 3.5-3.6 |
| | SCAN+$U_{Mn}$ | 9.45 | 9.48 | 9.47 | | 90 | | 0.19 | 3.8-3.9 |
| MnO$_2$ ($P4_2/mnm$) | Expt. | | 4.40 | 2.87 | | 90 | | 0.27-0.3 | 2.35 |
| | SCAN | | 4.38 | 2.85 | | 90 | | 0.0 | 2.62 |
| | SCAN+$U_{Mn}$ | | 4.40 | 2.88 | | 90 | | 0.64 | 2.84 |
| Mn$_3$O$_4$ ($I4_1/amd$) | Expt. | 5.75 | 6.22 | 5.75 | 117.5 | 90 | 117.5 | 2.3-2.5 | 4.34,[a] 3.25-3.64[b] |
| | SCAN | 5.75 | 6.18 | 5.75 | 117.4 | 90.5 | 117.4 | 0.68 | 4.37,[a] 3.63[b] |
| | SCAN+$U_{Mn}$ | 5.80 | 6.23 | 5.80 | 117.5 | 90.5 | 117.5 | 1.47 | 4.55,[a] 3.8[b] |
| FeO ($Fm\bar{3}m$) | Expt. | | 6.08 | | | 60 | | 2.4 | 3.32-4.2 |
| | SCAN | | 5.83 | 6.06 | 63.7 | 63.7 | 61.6 | 0.0 | 3.53 |
| | SCAN+$U_{Fe}$ | 6.09 | 6.16 | 6.09 | 59.9 | 61.2 | 59.9 | 0.71 | 3.7 |
| Fe$_2$O$_3$ ($R\bar{3}c$) | Expt. | | 5.04 | 27.54 | 90 | | 120 | 2.2 | 4.9 |
| | SCAN | | 5.03 | 27.47 | 90 | | 120 | 0.73 | 4.0 |
| | SCAN+$U_{Fe}$ | | 5.05 | 27.50 | 90 | | 120 | 1.98 | 4.3 |
| Fe$_3$O$_4$ ($Fd\bar{3}m$) | Expt. | | 8.39 | | | 90 | | 0.14 | 4.44,[c] 4.10[d] |
| | SCAN | | 8.34 | | | 90 | | 0.0 | 3.84,[e] 3.78[f] |
| | SCAN+$U_{Fe}$ | 8.41 | 8.42 | 8.43 | 89.8 | 89.9 | 89.8 | 0.6 | 4.2,[e] 3.7[f] |

[a]Mn$^{2+}$ (tetrahedral sites), [b]Mn$^{3+}$ (octahedral sites), [c]Tetrahedral sites, [d]Octahedral sites, [e]Fe$^{3+}$, and [f]Fe$^{2+}$ (tetrahedral + octahedral sites)

For most of the TMOs considered here, *e.g.*, CeO$_2$, Ce$_2$O$_3$, MnO$_2$, and Fe$_2$O$_3$, both DFT-SCAN and SCAN+$U$ lattice parameters are in fair agreement with experiments. For most of the oxides, DFT-SCAN tends to marginally underestimate the lattice constants (*i.e.*, over-binds) with respect to experimental values, with $CeO_2$ being a notable exception. Also, across systems, SCAN+$U$-predicted lattice constants are larger compared to DFT-SCAN (analogous to GGA+$U$ and DFT-GGA calculations[45,76,81]) and tend to be in slightly better agreement with experiments, with $CeO_2$, $Mn_2O_3$, and $Mn_3O_4$ being exceptions. DFT-SCAN spuriously underestimates some of the lattice constants, specifically $b$ in MnO and $a$ and $b$ in FeO, which is partially corrected for MnO by use of SCAN+$U_{Mn}$. The significant deviations for both DFT-SCAN and SCAN+$U_{Fe}$ predictions (versus experiments) in FeO can be attributed partly to the significant concentration of Fe-vacancies that tend to exist within the material at room temperature.[102] Interestingly, SCAN+$U_{Fe}$ captures the low-temperature cubic → orthorhombic transition in spinel-$Fe_3O_4$, unlike DFT-SCAN, signifying the importance of using a SCAN+$U$ framework for describing the right ground-state polymorph within TMOs.

As a ground-state theoretical framework, SCAN(+$U$) is not expected to precisely predict the band gaps of various structures, analogous to trends observed with GGA(+$U$) calculations.[37] SCAN(+$U$), however, has to qualitatively obtain the right ground-state electronic structure (*e.g.*, metallic versus semiconducting) if SCAN(+$U$) energies and structures are to be reliable. Qualitative trends in **Table 3** indicate significant discrepancies between DFT-SCAN electronic structures and experiments across several oxides, with DFT-SCAN systematically underestimating band gaps, consistent with prior observations in sulfides.[34] For example, DFT-SCAN predicts metallic behavior for $Ce_2O_3$, $Mn_2O_3$, $MnO_2$, FeO, and $Fe_3O_4$, in contrast to their observed semiconducting behavior at low temperatures.[73,96–99,101] The deviations in DFT-SCAN band gaps are particularly severe for $Ce_2O_3$, $Mn_2O_3$, and FeO, which exhibit reasonable band gaps experimentally (> 1 eV). We do not find semiconducting behavior for $MnO_2$ with SCAN (DOS plotted in **Figure S2** of the SI), unlike a previous report[35] where the authors used a fairly coarse $k$-point grid (at intervals of 0.25 Å$^{-1}$). We employed a significantly denser $k$-point mesh in our calculations (~ 0.03 Å$^{-1}$), which could explain the differences in the respective predictions.

Qualitative band-gap predictions indeed improve within SCAN+$U$, similar to improvements observed in GGA+$U$ calculations versus DFT-GGA.[36,37,40] For example, SCAN+$U$ correctly predicts a semiconducting electronic structure for $Ce_2O_3$, $Mn_2O_3$, $MnO_2$, FeO, and $Fe_3O_4$, unlike DFT-SCAN.

SCAN+$U$ band gaps also are in better agreement with experimental observations, although, as expected, SCAN+$U$ tends to underestimate (*e.g.*, $CeO_2$, $Ce_2O_3$, $MnO$, $Mn_2O_3$, $Mn_3O_4$, $FeO$, and $Fe_2O_3$) band gaps, being a ground-state theory. The occasional overestimate ($MnO_2$ or $Fe_3O_4$) is predicted for those with near-zero band gaps. Recall that such ground-state eigenvalue band gaps from DFT-SCAN/SCAN+$U$ do not correspond to measured optical or photoemission/inverse photoemission (quasiparticle) gaps and hence are not expected to yield a fair quantitative comparison.

In the case of magnetic moments on TM atoms ($m$), both DFT-SCAN and SCAN+$U$ qualitatively agree with experiments. Across all TMOs, SCAN+$U$ predicts higher absolute magnetic moments in comparison to SCAN, leading to quantitatively better agreement with experiments in a few cases (*e.g.*, $Ce_2O_3$, $MnO$, and $Fe_2O_3$). Notably, SCAN+$U$ leads to better charge localization on the Fe $d$ orbitals and the eventual ordering of $Fe^{2+}$ and $Fe^{3+}$ ions within $Fe_3O_4$, as signified by the distinct $m$ values (4.2 for $Fe^{3+}$ and 3.7 for $Fe^{2+}$), compared to DFT-SCAN ($m \sim 3.84$ and 3.78). The experimental magnetic moments in $Fe_3O_4$[89] represent the average Fe magnetic moments amongst the tetrahedral (occupied by $Fe^{3+}$, ~4.44) and octahedral (occupied by both $Fe^{2+}$ and $Fe^{3+}$, ~4.10) sites of the spinel structure. On the other hand, the SCAN+$U$ predictions correspond to individual $Fe^{2+}$ and $Fe^{3+}$ atoms irrespective of tetrahedral or octahedral occupancy, which explains the discrepancy observed between experiments and SCAN+$U$. In general, the higher absolute magnetic moments predicted by SCAN+$U$ (versus DFT-SCAN) is not surprising, because adding a $U$ facilitates electron localization on TM $d$ orbitals. However, the addition of $U$ in a few TMOs, such as $MnO_2$ and $Mn_3O_4$, leads to a worse agreement with experimental magnetic moments.

**Polymorph selection in Ce$_2$O$_3$**

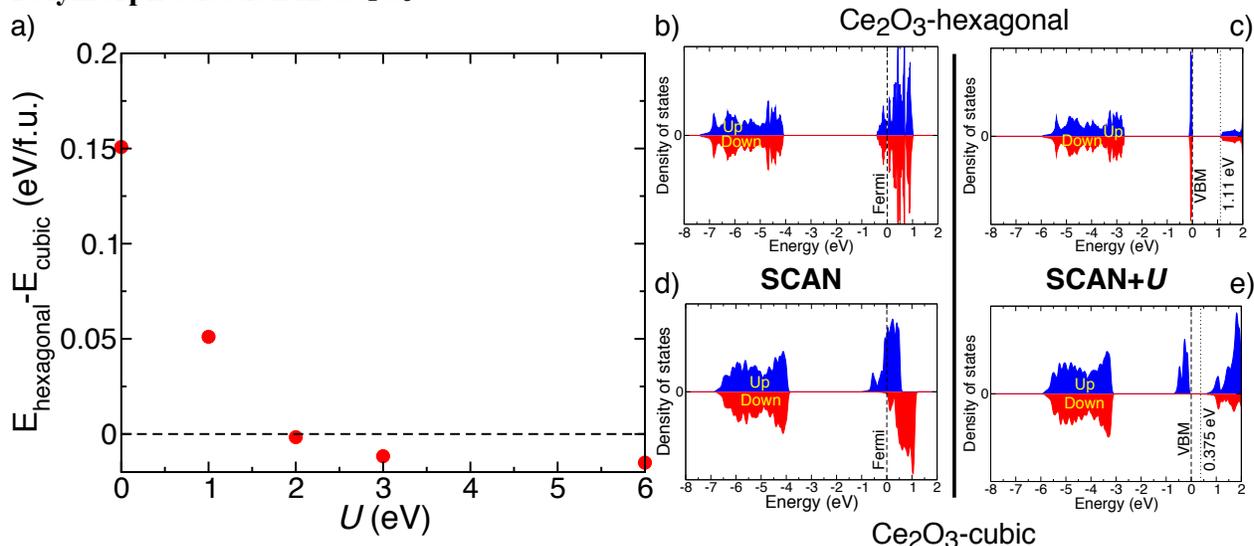

**Figure 4:** Stability of cubic-Ce$_2$O$_3$ (space group: $Ia\bar{3}$) versus hexagonal-Ce$_2$O$_3$ (space group: $P\bar{3}m1$) is plotted in panel a as a function of $U$ in SCAN+$U$. Total DOS of hexagonal-Ce$_2$O$_3$ (panels b and c) and cubic-Ce$_2$O$_3$ (panels d and e), as calculated via DFT-SCAN (panels b and d) and SCAN+$U$ (panels c and e), where $U_{Ce}$ = 2.0 eV. The zero on the energy scale in panels b – e are set either to the Fermi level (dashed lines in panels b and d) or to the valence band maximum (VBM, dashed lines in panels c and e). The dotted lines in panels c and e refer to the conduction band minimum (CBM), situated across the band gap. Blue (red) shaded regions correspond to the electronic up (down) spin states.

Experimentally, Ce$_2$O$_3$ can exhibit two distinct polymorphs: *i*) hexagonal (space group: $P\bar{3}m1$), which is the ground state and has been the subject of several theoretical and experimental studies;[45,49,80,81,94] and *ii*) cubic (space group: $Ia\bar{3}$), which is derived from an oxygen-deficient supercell of the fluorite-CeO$_2$ structure and is identical to the Bixbyite polymorph of Mn$_2$O$_3$.[103] Experimentally, cubic-Ce$_2$O$_3$ has been obtained via a deep reduction of fluorite-CeO$_2$.[103] However, there are no clear indications in literature of the energy difference between the two polymorphs. It is imperative, given the two polymorphs of Ce$_2$O$_3$, that any theoretical framework chosen to describe the energetics (or electronic structure), is able to identify the right ground state. **Figure 4a** consequently plots the energy difference between the hexagonal and cubic polymorphs of Ce$_2$O$_3$ across different $U$ values within the SCAN+$U$ framework.

DFT-SCAN ($U = 0$) wrongly predicts cubic-$Ce_2O_3$ to be the ground state (by ~0.15 eV/f.u. versus hexagonal-$Ce_2O_3$), while increasing $U$ values gradually increases the stability of hexagonal-$Ce_2O_3$. At $U = 2$ eV, the hexagonal polymorph is predicted to be the ground state (by ~1 meV/f.u.), with the stability of the hexagonal structure increasing up to ~15 meV/f.u. at $U = 6$ eV. Given that $U = 1.8$ eV minimizes the error between SCAN+$U$ calculations and experimental oxidation enthalpy for $Ce_2O_3 \rightarrow CeO_2$, and that a minimum of $U = 2$ eV is required to identify the right hexagonal ground state at $Ce_2O_3$, we define the optimal $U$ for Ce ($U_{Ce}$) to be 2 eV.

Panels **b**, **c**, **d**, and **e** in **Figure 4** plot the total DOS calculated in hexagonal-$Ce_2O_3$ (panels b and c) and in cubic-$Ce_2O_3$ (panels d and e). Panels b and d in **Figure 4** are calculated with DFT-SCAN, while panels c and e are calculated with SCAN+$U_{Ce}$. Shaded blue (red) components correspond to the electronic up (down) spin states. Dashed black lines in **Figure 4b** and **d** represent the Fermi level, while in panels **c** and **e** they represent the valence band maximum (VBM), with the energy scale (horizontal axis) set to zero either at the Fermi level or at the VBM. Also, the dotted black lines in **Figure 4c** and **e** signify the conduction band minimum (CBM), which are situated across the band gap.

Notably, DFT-SCAN incorrectly predicts metallic behavior for both the hexagonal (**Figure 4b**) and cubic (**Figure 4d**) polymorphs of $Ce_2O_3$. While band-gap measurements in hexagonal-$Ce_2O_3$ have reported an optical gap of ~2.4 eV,[94] such measurements do not exist for cubic-$Ce_2O_3$. However, measurements of electronic conductivity in reduced fluorite-$CeO_2$ (which is isostructural with cubic-$Ce_2O_3$) show thermally activated behavior,[104] which is typical of a semiconductor. Such discrepancies in the qualitative nature of the electronic structure using DFT-SCAN, *i.e.*, metallic versus semiconducting, introduces errors in the evaluation of the energies of the cubic and hexagonal polymorphs, leading to the prediction of a wrong ground state (**Figure 4a**). On the other hand, SCAN+$U_{Ce}$ calculations (**Figure 4c** and **e**) predict semiconducting behavior for both polymorphs of $Ce_2O_3$, with band (eigenvalue) gaps of ~1.11 eV (hexagonal) and ~0.375 eV (cubic). Although, as expected, SCAN+$U_{Ce}$ underestimates the band gap of hexagonal-$Ce_2O_3$ with respect to experiments, similar to trends observed in GGA+$U$ calculations, the qualitative description of a semiconducting electronic structure is correct for both the hexagonal and cubic polymorphs, resulting in the identification of the correct ground-state structure of $Ce_2O_3$.

**Magnetic configurations in $Mn_2O_3$**

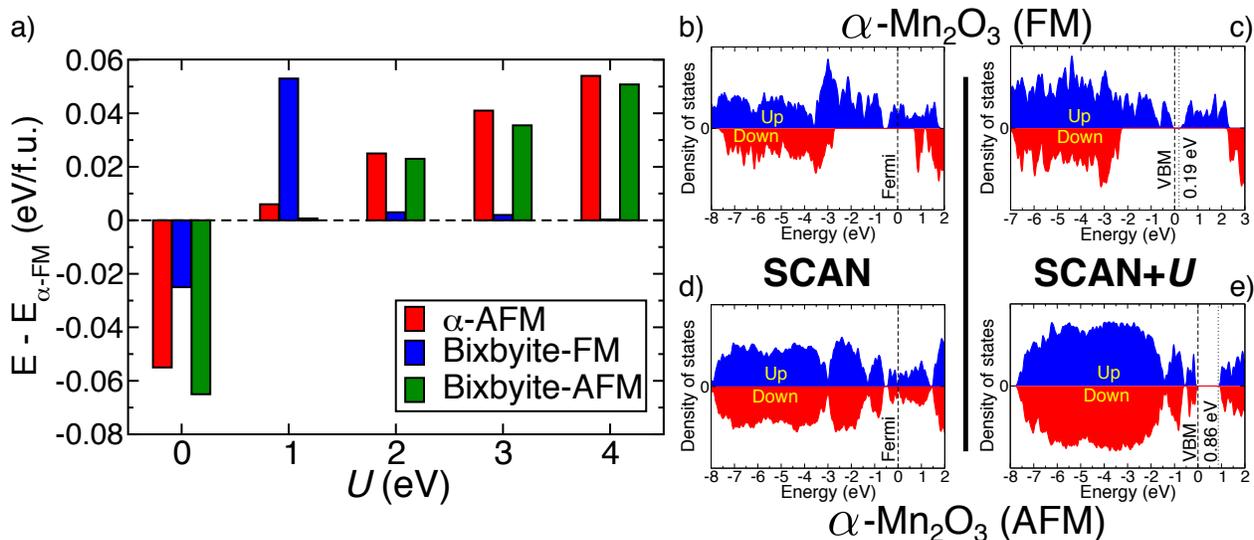

**Figure 5:** Stability of various magnetic and polymorphic configurations of $Mn_2O_3$ (panel a), namely, an AFM configuration of $\alpha$-$Mn_2O_3$ (space group: $Pbca$, red bars), the FM configuration of Bixbyite-$Mn_2O_3$ (space group: $Ia\bar{3}$, blue bars), and an AFM configuration of Bixbyite-$Mn_2O_3$ (green bars), are plotted as a function of $U$ within SCAN+$U$. The stabilities are referenced to FM $\alpha$-$Mn_2O_3$. Total DOS of FM (panels b and c) and AFM (panels d and e) configurations of $\alpha$-$Mn_2O_3$, as calculated via DFT-SCAN (panels b and d) and SCAN+$U$ (panels c and e), where $U_{Mn}$ = 2.7 eV. The zero on the energy scale in panels b – e is set either to the Fermi level (dashed lines in panels b and d) or to the VBM (dashed lines in panels c and e). The dotted lines in panels c and e refer to the CBM, situated across the band gap. Blue (red) shaded regions correspond to the electronic up (down) spin states.

While $Mn_2O_3$ exhibits an undistorted cubic structure (Bixbyite polymorph, space group: $Ia\bar{3}$) at temperatures above 302 K, the compound undergoes a cubic → orthorhombic (space group: $Pbca$, referred to as $\alpha$-$Mn_2O_3$) transition at lower temperatures, resulting in ~0.8% deviation away from cubic symmetry.[77] Additionally, $\alpha$-$Mn_2O_3$ is known to undergo a paramagnetic → AFM transition when cooled below ~90 K.[105] Thus, the AFM $\alpha$-$Mn_2O_3$ is the true ground-state configuration. **Figure 5a** plots the energies of three distinct magnetic configurations, namely, AFM $\alpha$-$Mn_2O_3$ (red bars), FM Bixbyite-$Mn_2O_3$ (blue bars), and AFM Bixbyite-$Mn_2O_3$ (green bars), at different $U$ values. All energies in **Figure 5a** are referenced to FM $\alpha$-$Mn_2O_3$ (dashed black line).

Without any $U$ added to Mn atoms, DFT-SCAN predicts the AFM Bixbyite configuration to be the ground state, which is more stable than FM $\alpha$-$Mn_2O_3$ and FM Bixbyite by ~0.065 eV/f.u. and ~0.01 eV/f.u., respectively. DFT-SCAN thus incorrectly predicts the ground-state polymorph for $Mn_2O_3$, similar to $Ce_2O_3$ (**Figure 4**). In contrast, SCAN+$U$ predicts $\alpha$-$Mn_2O_3$ to be the ground-state polymorph across all $U$ values, consistent with experiments. However, SCAN+$U$ incorrectly predicts the magnetic ground-state configuration for $\alpha$-$Mn_2O_3$, with the FM configuration more stable than AFM by ~6, 25, 41, and 56 meV/f.u. at $U$ values of 1, 2, 3, and 4 eV, respectively. Specifically, at the optimal $U$ value for Mn ($U_{Mn}$ = 2.7 eV, **Figure 3b**), the FM $\alpha$-$Mn_2O_3$ is predicted to be more stable than the corresponding AFM configuration by ~36 meV/f.u. The discrepancy in the magnetic ground-state configuration predicted by SCAN+$U$ may be attributed to the specific AFM ordering used in our calculations, which was originally proposed by Regulski *et al.*[77] Note that the ground-state AFM ordering of $\alpha$-$Mn_2O_3$ is still debated.[76,77] Also, the small difference in energy between the FM and AFM $\alpha$-$Mn_2O_3$ configurations (25-40 meV/f.u.), between $U$ values of 2 and 3 eV, would not change the optimal $U$ value significantly, *i.e.*, $U_{Mn}$ varies by < 0.01 eV (**Figure 3b**). This suggests that the magnetic configuration of $\alpha$-$Mn_2O_3$ plays an insignificant role in the overall redox energetics of Mn-oxides. At U $\geq$ 2 eV, Bixbyite-$Mn_2O_3$ initialized as either FM or AFM interestingly exhibits the cubic $\rightarrow$ orthorhombic distortion during our structural relaxation calculations, as indicated by the similarity in energies ($\pm$ 5 meV/f.u.) of FM Bixbyite with FM $\alpha$-$Mn_2O_3$ and AFM Bixbyite with AFM $\alpha$-$Mn_2O_3$, respectively. We do not observe any cubic $\rightarrow$ orthorhombic distortion in our DFT-SCAN or SCAN+$U$ calculations of FM/AFM Bixbyite.

Panels **b** and **c** (**d** and **e**) of **Figure 5** plot the total DOS for FM (AFM) $\alpha$-$Mn_2O_3$, calculated using DFT-SCAN (panels **b** and **d**) and SCAN+$U_{Mn}$ (panels **c** and **e**). Analogous to $Ce_2O_3$, DFT-SCAN incorrectly predicts a metallic ground-state electronic configuration for $\alpha$-$Mn_2O_3$ in both the FM and AFM configurations (**Figure 5b** and **d**). Experimentally, $\alpha$-$Mn_2O_3$ exhibits a band gap of 1.2-1.3 eV,[98,99] which is indicative of a distinct semiconducting behavior. Adding a $U$ on the Mn atoms does correct the qualitative description of the electronic ground state, with a predicted band (eigenvalue) gap of ~0.19 and 0.86 eV for the FM and AFM configurations, respectively, at $U_{Mn}$. However, the SCAN+$U_{Mn}$ band gap for FM $\alpha$-Mn$_2$O$_3$ is quite small in comparison to AFM $\alpha$-Mn$_2$O$_3$ and experiments. Similarly, DOS calculations in FM and AFM Bixbyite-Mn$_2$O$_3$ (not shown) display (near-)metallic and semiconducting natures in both the DFT-SCAN and SCAN+$U_{Mn}$ calculations, respectively. Specifically, band gaps for AFM Bixbyite are ~0.11 and 1.08 eV in DFT-SCAN and SCAN+$U_{Mn}$ frameworks, respectively. On the other hand, FM Bixbyite is predicted to be metallic by both DFT-SCAN and SCAN+$U_{Mn}$. Thus, even though the FM and AFM Bixbyite-Mn$_2$O$_3$ exhibit cubic → orthorhombic distortions in our SCAN+$U_{Mn}$ calculations, they relax to different electronic ground states.

**Electronic structure in Fe$_3$O$_4$**

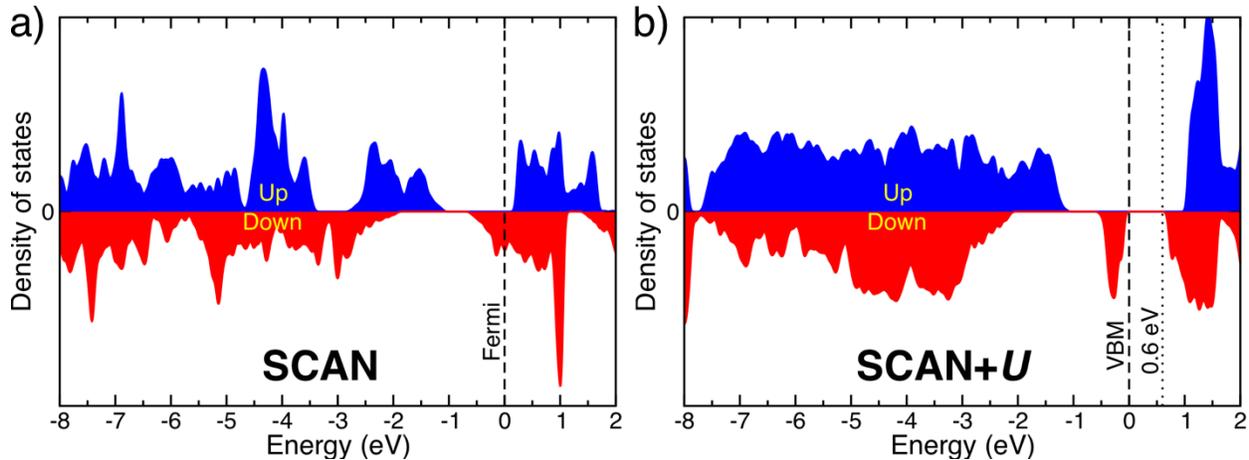

**Figure 6:** Total DOS in ferrimagnetic spinel-Fe$_3$O$_4$, as calculated by DFT-SCAN (panel a) and SCAN+$U$ (panel b), where $U_{Fe}$ = 3.1 eV. Shaded blue (red) regions correspond to up (down) electronic spin states. Dashed lines in both panels indicates the zero on the energy scale, which is set to the Fermi level in panel a and to the VBM in panel b. The dotted line in panel b indicates the CBM, which is situated across a ~0.6 eV band gap.

Spinel-Fe$_3$O$_4$ displays moderate electrical conductivity at room temperature, which has been attributed to electron delocalization across Fe$^{2+}$ and Fe$^{3+}$ octahedral sites, resulting from inversion in the spinel structure.[69,72] However, at temperatures below 120 K (the Verwey transition temperature), charge ordering of the $Fe^{2+}$ and $Fe^{3+}$ takes place on the octahedral sites, resulting in a small band gap (~0.14 eV[73]) and a ferrimagnetic magnetic ground state. The charge ordering also leads to a slight cubic → orthorhombic distortion of the spinel structure, although some studies instead have reported the distortion to be monoclinic.[69] It therefore is imperative that theoretical calculations capture *i*) the distortion away from the cubic structure; *ii*) the opening of the band gap at low temperatures; and *iii*) the ferrimagnetic ground-state configuration. Notably, both DFT-SCAN and SCAN+$U_{Fe}$ calculations predict a ferrimagnetic ground-state configuration (pictorially represented in **Figure 1**), although DFT-SCAN does not capture any relaxations away from the cubic symmetry of the spinel structure (**Table 1**). Analogous to $Ce_2O_3$ (**Figure 4**) and $Mn_2O_3$ (**Figure 5**), we observe that DFT-SCAN incorrectly predicts a metallic ground-state electronic configuration (see total DOS calculations plotted in **Figure 6a**) for ferrimagnetic spinel-$Fe_3O_4$, which may have caused the cubic symmetry to be conserved during DFT-SCAN calculations. In the case of SCAN+$U_{Fe}$, the band gap is overestimated, *i.e.*, ~0.6 eV versus 0.14 eV experimentally (**Figure 6b**). However, SCAN+$U_{Fe}$ does predict a ferrimagnetic ground state, a non-negligible band gap, and a distortion away from the cubic spinel symmetry, consistent with experimental observations and signifying a satisfactory description of the ground-state configuration of $Fe_3O_4$.

## Discussion

In this work, we evaluate the oxidation energetics, lattice parameters, band gaps, and magnetism in Ce-, Mn-, and Fe-oxides (**Figure 1**), which constitute important TMO materials for STC applications, to examine the applicability of DFT-SCAN for describing redox energetics amongst TMOs. We find that DFT-SCAN exhibits excellent agreement with experimental formation energies of main group binary oxides (**Figure 2**) and measured bond dissociation energy of the $O_2$ molecule. Given the significant overestimation of oxidation enthalpies within TMOs by DFT-SCAN, we calculated optimal $U$ values based on oxidation energies of available binary oxides for Ce (2 eV), Mn (2.7 eV), and Fe (3.1 eV, **Figure 3**) cations. In addition to oxidation energetics, we also benchmarked the lattice parameters, band gaps, and TM magnetic moments of all of the oxides considered against experiments (**Table 1**) and find that DFT-SCAN severely underestimates band gaps across all TMOs.

The inaccurate description of the electronic behavior by DFT-SCAN leads to incorrect predictions of ground-state configurations in $Ce_2O_3$ (**Figure 4**), $Mn_2O_3$ (**Figure 5**), and $Fe_3O_4$ (**Figure 6**). Specifically, DFT-SCAN incorrectly predicts the cubic polymorph of $Ce_2O_3$ (versus hexagonal), the AFM Bixbyite configuration of $Mn_2O_3$ (versus AFM $\alpha$-$Mn_2O_3$), and the metallic cubic spinel-$Fe_3O_4$ (versus semi-conducting distorted spinel-$Fe_3O_4$) to be the ground states. Adding a $U$ to SCAN indeed overcomes some of the limitations of DFT-SCAN, namely, prediction of the right ground-state structures of $Ce_2O_3$ (hexagonal), $Mn_2O_3$ ($\alpha$), and $Fe_3O_4$ (distorted spinel). The major difference between SCAN+$U$ and DFT-SCAN is the qualitative agreement of the former with experiments regarding the semiconducting behavior of all of the oxides considered. In the specific case of spinel-$Fe_3O_4$, the superior charge localization in $Fe_3O_4$ within SCAN+$U$ ($m_{Fe}$ in **Table 1**) may have facilitated the cubic → orthorhombic distortion during the structure relaxation, unlike DFT-SCAN. A superior qualitative description of the electronic structure therefore can lead to a better quantitative description of the energetics.

Notably, SCAN+$U$ also underestimates the band gaps of most oxides (**Table 1**) with wider band gaps (> 1 eV), analogous to trends observed in GGA+$U$ calculations. This is expected when using a ground-state theory to predict excited-state properties. However, for small band-gap semiconductors (< 1 eV), such as $MnO_2$ and $Fe_3O_4$, SCAN+$U$ calculations do overestimate the experimental band gaps. Significantly, SCAN+$U$ does not predict the precise ground-state magnetic configuration in $\alpha$-$Mn_2O_3$, which may be attributed to the AFM model used in this work.[77] Note that the erroneous magnetic ground state of $\alpha$-$Mn_2O_3$ does not significantly affect the optimal $U$ value determined and hence has a negligible impact on redox energetics involving $Mn^{+2/+3/+4}$ ions. This is expected, since the energetic scale of magnetic interactions in most solids is typically orders of magnitude lower than redox energetics.[106] Nevertheless, the SCAN+$U$ framework needs to be carefully benchmarked for each TMO before being used to predict material properties.

Interestingly, the optimal $U$ values determined in this work are significantly lower than the corresponding $U$ values determined with a GGA XC functional. For example, previous fits of $U$ on Ce-oxides prescribe a value between 2-3 eV,[45,81] while we evaluate $U_{Ce}$ = 2 eV for SCAN+$U$. Similarly, $U$ = 3.9-4 eV was determined based on oxidation energetics in Mn-oxides with GGA+$U$,[42,50] while we estimate a lower $U_{Mn}$ = 2.7 eV. Also, *ab initio* evaluations of $U$ for $Fe^{2+}$ and $Fe^{3+}$ with GGA are 3.7 and 4.2 eV, respectively,[40,41] which are much higher than $U_{Fe}$ = 3.1 eV determined in this work. Finally, $U$ calculated from linear response theory,[107] based on GGA calculations, also yielded significantly higher values than those determined in our work, namely $U_{Ce}$ = 4.5 eV,[81] $U_{Mn}$ = 3.92 ($Mn^{2+}$), 4.64-5.09 eV ($Mn^{3+}$), 5.04 eV ($Mn^{4+}$),[43] and $U_{Fe}$ = 3.71 eV ($Fe^{2+}$), 4.9 eV ($Fe^{3+}$).[43] The magnitude of the $U$ value required should be dependent on the accuracy of the electronic exchange interactions that are captured by the XC functional, *i.e.*, the more accurate the XC functional, the lower the $U$ value required. Thus, the $U$ values required with SCAN for other TM systems are likely to be significantly lower than the corresponding $U$ values determined for GGA+$U$ (or LDA+$U$) calculations, due to a better capture of electronic exchange interactions by SCAN versus GGA.

**Conclusions**

Solar thermochemical (STC) technology could be an important contributor to the generation of reusable fuels using renewable energy sources, where there is an urgent need of innovation in the material substrates used, motivating quantum-mechanics-based screenings of STC materials. Notably, any quantum-mechanics framework, such as DFT calculations, used for materials design must correctly describe the energetics and electronic structure changes of the redox reactions involved. We therefore benchmarked DFT-SCAN oxidation enthalpies, lattice parameters, and band gaps of binary Ce-, Mn-, and Fe-oxides, which are important current ingredients for STC applications,[3,4,6] in addition to evaluating formation enthalpies of main group oxides. Based on the excellent agreement between experimental and DFT-SCAN oxide formation enthalpies of main group elements, and for the $O_2$ bond dissociation energy, we conclude that DFT-SCAN does not over-bind the $O_2$ molecule, unlike DFT-LDA and DFT-GGA. However, the SCAN+$U$ framework was required to accurately describe the oxidation energetics of binary TMOs, with DFT-SCAN predictions significantly overestimating (*i.e.*, yielding too negative) oxidation enthalpies compared to experiments. Significantly, DFT-SCAN erroneously predicted the qualitative electronic structure of several TMOs considered in this work, leading to wrong polymorphs being predicted as ground states ($Ce_2O_3$, $Mn_2O_3$, and $Fe_3O_4$). Adding a $U$ on the TM centers mitigated the shortcomings of DFT-SCAN, with qualitative agreements with experiments on the electronic behavior and, subsequently, the ground-state polymorphs. SCAN+$U$ therefore yielded better ground-state energies and lattice parameters via accurate descriptions of the electronic structure. In the case of magnetic moments on TM atoms, both DFT-SCAN and SCAN+$U$ calculations qualitatively agreed with experiments. Interestingly, we found the optimal $U$ values determined for SCAN+$U$ calculations to be systematically lower than the corresponding $U$ values frequently employed in GGA+$U$ calculations, signifying the improved electronic exchange description of SCAN versus GGA. Finally, we recommend using the SCAN+$U$ functional for describing redox reactions involving other TMOs and sulfides, with the value of $U$ either determined via a rigorous benchmarking with experimental data as described in this work or using *ab initio* methods as elaborated elsewhere.[40,41]

## Acknowledgments

E.A.C. thanks the U.S. Department of Energy, Office of Energy Efficiency and Renewable Energy under Award No. DE-EE0008090 for funding this project. The authors thank Ms. Nari L. Baughman for a careful reading of the manuscript.